\documentclass{emulateapj}
\usepackage{epsfig}

\newcommand{\etal}{\mbox{et al.}}

\newcommand{\ergs}{erg s$^{-1}$}

\newcommand{\degree}{$^\circ$}
\newcommand{\msun}{$M_{\odot}$}

\newcommand{\spitzer}{{\it Spitzer}}

\shortauthors{Muno \& Mauerhan}
\shorttitle{Mid-Infrared Excesses from LMXBs}

\begin{document}

\title{Mid-Infrared Emission from Dust around Quiescent Low-Mass X-ray Binaries}

\author{
Michael P. Muno\altaffilmark{1} and Jon Mauerhan}
\affil{Department of Physics and Astronomy, University of California, Los
Angeles}

\altaffiltext{1}{Hubble Fellow}

\begin{abstract}
We report the discovery of excess 4.5 and 8$\mu$m emission from three 
quiescent black hole low-mass X-ray binaries, A 0620$-$00, GS 2023+338, and 
XTE J1118+480, and the lack of similar excess emission from Cen X-4. 
The mid-infrared emission from GS 2023+338 probably originates
in the accretion disk. However, the excess emission from A 0620$-$00 and 
XTE J1118+480 is brighter and peaks at longer wavelengths, and most likely
originates from circumbinary dust that is heated by the light of 
the secondary star. For these two sources, 
we find that the inner edges of the dust distributions lie near 1.7 times 
the binary separations, which are the minimum radie at 
which circumbinary disks would be stable against tidal disruption. The 
excesses are weak at 24 $\mu$m, which implies that 
the dust does not extend beyond about 3 times the binary separations. The 
total masses of circumbinary material are between $10^{22}$ and $10^{24}$ g.
The material could be the remains of fall-back disks produced in supernovae,
or material from the companions injected into circumbinary orbits during 
mass transfer.
\end{abstract}


\section{Introduction}

One of the main limiting factors in determining the masses of the 
compact objects in low-mass X-ray binaries 
is uncertainty in the fraction of the infrared light that is produced
by the companion star (e.g., Hynes, Robinson, \& Bitner 2005). When 
LMXBs are accreting at low rates, the optical and infrared light from 
these systems is dominated by the companion stars, so their mass
functions can be measured to high accuracy based on the Doppler motion of the 
companion \citep[e.g.,][]{mr06}. The inclination then must be 
constrained by modeling the modulations in the optical and infrared 
light curves that are produced by the varying aspect 
of the distorted, Roche-lobe-filling companion. Generally, the contribution
of the accretion disk to the optical and near-infrared emission is uncertain, 
and varying the fraction of the light that is assumed to be produced by 
the accretion disk can lead to differences of a factor of 2 in the 
derived mass of a compact object (e.g., Gelino, Harrison, \& Orosz 2001).
Fortunately, the broad-band spectrum of a multi-temperature accretion
disk is significantly flatter than that of a stellar photosphere, so 
mid-infrared
observations could constrain the relative contributions of the two 
components.

However, the spectra of LMXBs in the mid-infrared have not been well 
studied, and several indirect lines of evidence suggest that these
systems might contain circumbinary material that could emit in the 
mid-infrared.
gFirst, some of the white 
dwarf analogs to LMXBs, cataclysmic variables (CVs), exhibit spectral 
features that lie at the mean radial velocities of the
systems \citep{ss94} and excess mid-infrared emission \citep{how06} 
that could be interpreted as arising in circumbinary 
material \citep[see also][]{taam03,dub04}.
Second, the supernovae that produced the compact objects could
have left fall-back disks around the binaries. Indeed, the first fall-back disk
has recently been found around a young, highly-magnetized neutron star
(Wang, Chakrabarty, \& Kaplan 2006)\nocite{wck06}. 
Third, the planets around the isolated millisecond 
pulsar PSR 1257$+$12 \citep{wf92} could not have survived the 
supernova that produced the neutron star, and must have formed 
afterward \citep[see, e.g.,][]{mh01}. 
Millisecond pulsars are usually assumed to have been spun up by accretion 
as LMXBs, in which case planets could form from material present during 
the binary phase. Therefore,  
to search for evidence of circumbinary material, we have observed four 
nearby, quiescent LMXBs with the {\it Spitzer} Space Telescope.
 
\section{Observations and Data Analysis}

We chose the LMXBs in our sample to be detectable
with \spitzer\ if they contained optically-thick circumbinary disks 
passively illuminated by the mass donor stars 
\citep[see][and below]{jur03,taam03}.
Based on a simple model, we chose sources with: 
(1) $K$ magnitudes brighter than 17, 
(2) locations more than $50^\circ$ in projection from the Galactic center, and
(3) no 2MASS sources within 5\arcsec\ that were brighter than 
our targets.
In Table \ref{tab:targets}, we list the positions, the orbital periods,
estimates of the primary masses, the spectral types of the companions, and 
the quiescent $K$ magnitudes of the four systems in our sample 
(see the table notes for references).

\begin{figure}
\centerline{\psfig{file=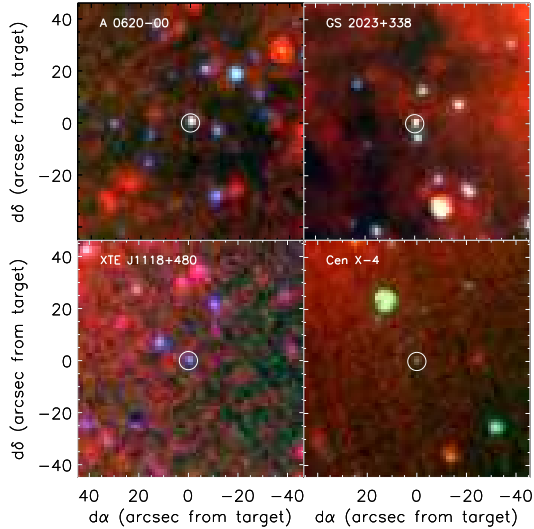,width=0.9\linewidth}}
\caption{
Mid-infrared images of the LMXBs in this survey. The red 
channel is 24 $\mu$m, the green 8.0 $\mu$m, and the blue
4.5 $\mu$m. The circles denote the positions of the optical
counterparts. 
}
\label{fig:img}
\end{figure}

\begin{figure}
\centerline{\psfig{file=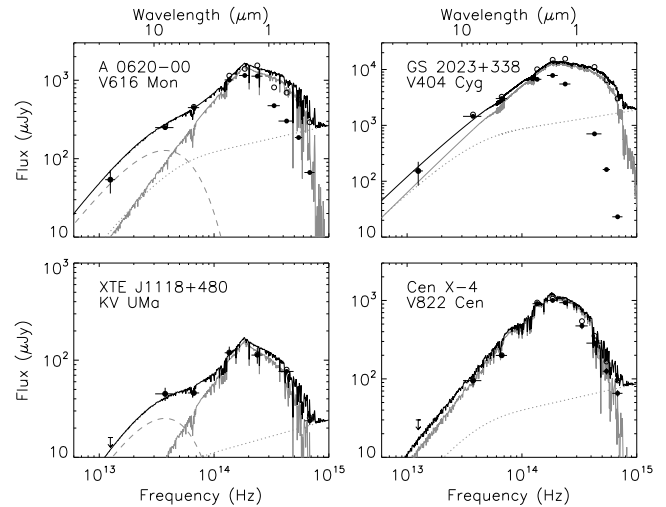,width=0.8\linewidth,angle=90}}
\caption{
The spectra of the LMXBs in our sample. 
The filled circles are the observed
fluxes in each band, and the open circles are the de-reddened fluxes 
(see notes to Tab. 1 for references). The upper limits on the 
24$\mu$m fluxes for XTE J1118+480 and Cen X-4 are indicated with arrows.
The uncertainties in the observed fluxes are also indicated, but are generally 
smaller than the symbol sizes. 
The solid grey lines are synthetic stellar spectra for the spectral
type in Table 1. The dotted grey lines are model spectra of viscously-heated
accretion disks, with accretion rates of $3\times10^{14}$ g s$^{-1}$ for
A 0620$-$00, $7\times10^{16}$ g s$^{-1}$ for GS 2023+338, 
$4\times10^{13}$ g s$^{-1}$ for XTE J1118+480, and 
$4\times10^{14}$ g s$^{-1}$ for Cen X-4. The grey dashed lines
illustrate the spectra of blackbodies used in the text to estimate the 
solid angles of the material responsible for the mid-infrared excesses from
A 0620$-$00 and XTE J1118+480. The black lines are the sum of these model
components, normalized to match the de-reddened $K$ flux.
}
\label{fig:sed}
\end{figure}

Our measurements were taken in the 4.5 and 8.0 $\mu$m bands with 
Infrared Array Camera (IRAC), and in the 24 $\mu$m band with the 
Multiband Imaging Photometer (MIPS; Tab.~\ref{tab:fluxes}). We used the 
post-basic-calibration data provided by the \spitzer\ Science Center
(SSC) for most of our analysis. However, the MIPS image of A 0620$-$00 
contained latent features with a low spatial frequency and a $\sim$2\% 
amplitude that were left by a previous observation of a bright, extended
source. We corrected the image by creating a flat field from the median
of the individual dithered images, dividing each snapshot by the flat, and
re-creating the mosaicked image using the script provided by the 
SSC.\footnote{http://ssc.spitzer.caltech.edu/postbcd/}
Three-color images centered on each LMXB are displayed in Figure~\ref{fig:img}.
Each target is detected at 4.5 and 8.0 $\mu$m. Only A 0620$-$00 and 
GS 2023+338 are also detected at 24 $\mu$m. 
We computed the IRAC fluxes of each source using the 
point-spread-function-fitting routine {\tt APEX} from the SSC, and the MIPS 
fluxes and upper limits using aperature photometry (Tab.~\ref{tab:fluxes} and 
Fig. \ref{fig:sed}).

\begin{deluxetable*}{llcccccccc}[htp]
\tablecolumns{10}
\tablewidth{0pc}
\tablecaption{Target Quiescent LMXBs\label{tab:targets}}
\tablehead{
\colhead{Source} & \colhead{Optical} & \colhead{RA} & \colhead{DEC} & 
\colhead{$D$} & \colhead{$P_{\rm orb}$} & \colhead{$M_1$} & 
\colhead{Sp.\ Type} & \colhead{$A_V$} & 
\colhead{$K$} \\
\colhead{} & \colhead{} & \multicolumn{2}{c}{(J2000, degrees)} & 
\colhead{(kpc)} & 
\colhead{(hr)} & \colhead{\msun} & \colhead{} & \colhead{} & 
\colhead{}
} 
\startdata
A 0620$-$00 & V616 Mon & 95.68561 & $-0.345628$ & 
  1.2$\pm$0.1 & 7.8 & 8.7--12.9 & 
  K4V & 1.2 & 
  14.55(6) \\
GS 2023+338 & V404 Cyg & 306.01594 & $+33.86728$ & 
  2.2--3.7 & 155.3 & 10.1--13.4 & 
  K0III & 4.0 & 
  12.50(5) \\
XTE J1118+480 & KV UMa & 169.54498 & $+48.03678$ & 
  1.8$\pm$0.5 & 4.1 & 6.5--7.2 & 
  K6V & 0.06 & 
  16.9(2) \\
Cen X-4 & V822 Cen & 224.59135 & $-31.66872$ & 
  1.2 & 15.1 & 1.4\tablenotemark{a} & 
  K5V & 0.3 & 
  14.66(8)
\enddata
\tablenotetext{a}{We assumed this mass for the neutron star based on 
those measured for binary radio pulsars \citep{tc99}.}
\tablecomments{The positions are taken from 2MASS, and the distances,
orbital periods, and mass function from \citet{mr06}. For the remaining
data, the references are --- A 0620$-$00: \citet{gho01} for all photometry
and the spectral type --- GS 2023+338: \citet{cas03} for all photometry
and the spectral type --- XTE J1118+480: spectral type from \citet{tor04};
extinction from \citet{mcc03}; J and K from \citet{mik05}; 
B from USNO; and R from \citet{zur02} --- 
Cen X-4: spectral type from Shahbaz, Naylor, \& Charles (1993);
\citet[][]{tor02}; J, H, K from 2MASS;
other bands from \citet{snc93}.
}
\end{deluxetable*}

\begin{deluxetable*}{lcccccccccc}[htb]
\tabletypesize{\scriptsize}
\tablecolumns{11}
\tablewidth{0pc}
\tablecaption{Spitzer Measurements of Quiescent LMXBs\label{tab:fluxes}}
\tablehead{
\colhead{} & \multicolumn{6}{c}{IRAC} & \multicolumn{4}{c}{MIPS} \\
\colhead{Source} & \colhead{Date IRAC} & 
\colhead{$T_{\rm IRAC}$} & \colhead{$S_{4.5}$} & \colhead{$\Delta S_{4.5}$} &
\colhead{$S_{8.0}$} & \colhead{$\Delta S_{8.0}$} &
\colhead{Date MIPS} &
\colhead{$T_{\rm MIPS}$} & \colhead{$S_{24}$} & \colhead{$\Delta S_{24}$} \\
\colhead{} & \colhead{} & 
\colhead{(s)} & \colhead{($\mu$Jy)} & \colhead{($\mu$Jy)} &
\colhead{($\mu$Jy)} & \colhead{($\mu$Jy)} &
\colhead{} &
\colhead{(s)} & \colhead{($\mu$Jy)} & \colhead{($\mu$Jy)}
} 
\startdata
A 0620$-$00 & 2005 Mar 25 & 400 & 448(13) & 194 & 249(10) & 149 &
  2005 Mar 06 & 180 & 54(18) & 43 \\
GS 2023+338 & 2004 Oct 09 & 36 & 3020(90) & 670 & 1450(40) & 500 &
  2004 Oct 16 & 30 & 153(70) & 46 \\
XTE J1118+480 & 2004 Nov 21 & 400 & 46(1) & 17 & 45(7) & 34 &
  2005 May 13 & 240 & $<$16 & $<$16 \\
Cen X-4 & 2004 Aug 12 & 300 & 199(6) & $-$90 & 95(17) & $-$14 & 
  2005 Aug 28 & 150 & $<$30 & $<$30 
\enddata
\tablecomments{We list the date and duration of each observation (the
4.5 and 8 $\mu$m data were obtained in parallel), the fluxes measured 
($S_\lambda$, where $\lambda$ is the central wavelength of each band), 
and estimates of the minimum amount of excess mid-infrared
emission above that expected from the photospheres of the companions
($\Delta S_\lambda$; see text). The quoted uncertainties are the 
larger of the standard deviation of the fluxes in the multiple frames
we obtained, or the systematic uncertainty in the absolute calibration
(3\% for IRAC, and 5\% for MIPS). The negative values for Cen X-4 reflect the
fact that extrapolating a $\lambda^2$ spectrum expected for the Rayleigh-Jeans
tail of the photosphere from the $K$-band flux predicts more
mid-infrared flux than is observed; this is likely because the infrared flux
is variable.}
\end{deluxetable*}

To understand the origin of the mid-infrared emission, in Figure \ref{fig:sed}
we plot for each LMXB the observed and de-reddened fluxes in the 
infrared and optical bands (from Tab.~\ref{tab:fluxes} and the references 
in Tab.~\ref{tab:targets}).
The optical and near-infrared intensities
of quiescent LMXBs often vary by several tenths of a magnitude on 
time scales of years \citep[not counting outbursts;][]{che89,gho01}, 
presumably because of changes in the accretion flow.
None of the fluxes were obtained simultaneously, so when comparing the 
flux measurements in Figure~\ref{fig:sed} one can only expect them to 
be self-consistent to within $\approx$30\%. 
To estimate the contributions of the companions to the spectra, we 
have obtained
model stellar spectra computed with the next-generation 
Phoenix code \citep[kindly provided by T. Barman;][]{hlb01}
that correspond to the temperatures and radii of the
spectral types in Table~\ref{tab:targets}. 
We have estimated the 
contribution of the accretion disk using a standard \citet{ss73}
model, assuming the disk extends out to the tidal truncation 
radius ($\la$60\% of the binary separation; Frank, King, \& Raine 1992),
is inclined by 60\degree\ to our line of sight, and has the largest
accretion rate that is consistent with the optical photometry (see 
the caption of Fig.~\ref{fig:sed} for values).
We have confirmed that our conclusions are robust against changing the 
assumed spectral types of the companions within the values reported in 
the literature, and against varying the parameters of the accretion disk so
long as the models are consistent with the optical photometry.

We find that the spectrum of Cen X-4 is consistent with that expected
for the companion star, with a possible contribution at long wavelengths
from the accretion disk. There are clear excesses of mid-infrared flux 
above that expected from the companions of A 0620-00, GS 2023+338, and
XTE J1118+480. 
In order to conservatively estimate the amounts of the mid-infrared fluxes that
are not produced by the companions, we computed the maximum contributions 
of the companions' photospheres by normalizing the model stellar spectra 
to match the de-reddened $K$ fluxes, and reported the difference between the 
predicted and observed mid-infrared fluxes as the 
excesses in Table \ref{tab:fluxes} 
($\Delta S_\lambda$).

\section{Discussion}

Several LMXBs have now been detected in outburst in the mid-infrared
\citep[e.g., Smith, Beall, \& Swain 1990;][]{vp90, hg91, hg94, mig06}, 
but these are the first detections of quiescent systems. In outburst, the 
mid-infrared emission could originate from an accretion disk in which 
$\dot{M} > 10^{18}$ g s$^{-1}$ \citep{sbs90}, 
from free-free emission in a strong wind driven from the accretion disk
\citep{vp90}, or from radio jets
with flat spectra extending to infrared wavelengths 
\citep{fen01,hom05,mig06}. 
We use these hypotheses as a starting point for understanding the 
mid-infrared emission from our sample of quiescent LMXBs. 

The mid-infrared excess from GS 2023+338 could originate in an accretion
disk, because the 2.2--24 $\mu$m fluxes in 
Figure~\ref{fig:sed} follow 
the $v^{-2}$ law that one 
would expect from the Rayleigh-Jeans tail of a black body. We find that
our model for a viscously-heated 
accretion disk \citep{ss73} from \S2 adequately describes the excess
mid-infrared flux from GS 2023+338 in Figure~\ref{fig:sed}.

In contrast, the 4--24 $\mu$m fluxes from A 0620$-$00 and XTE J1118+480 do
not follow the $\nu^{-2}$ law expected for the Rayleigh-Jeans tail of a 
blackbody, because there is too much flux at 8 $\mu$m. A blackbody spectrum
that peaks near 8 $\mu$m would have a temperature of only 640 K. For these 
two sources, we plot blackbody spectra that match the peak of the 
mid-infrared excesses ($\Delta S_{8.0}$) 
in Figure~\ref{fig:sed}. The sum of the
spectra of stellar photospheres and optically-thick 
blackbodies match the observed mid-infrared fluxes well. However, the 
inferred emitting areas are large compared to the binary separations.
For a Planck spectrum, the 
solid angles of the emitting regions correspond to circular radii of  
\begin{equation} 
R = 4\times10^{11} \left( \frac{\Delta S_{8 \mu{\rm m}}}{100 \mu{\rm Jy}} \right)^{1/2} \left( \frac{D}{{\rm kpc}} \right)~{\rm cm},
\end{equation}
where $D$ is the distance to the source. 
In contrast, the disks will be 
contained within $\la$60\% of the binary separations $a$, which are given by:
\begin{equation}
a = 3\times10^{11} M^{1/3} (1+q)^{1/3} P_{\rm day}^{2/3}~{\rm cm},
\label{eq:orbit_sep}
\end{equation}
where $P_{\rm day}$ is the orbital period in days, $M$ is the 
mass of the accretor in solar masses, and $q$ is the mass ratio of the
binary \citep[e.g.,][]{fkr92}. 
Given the fluxes in Table \ref{tab:fluxes}, 
we find that the excess mid-infrared 
emission originates from regions $\approx$2 times larger than the orbital 
separations of GS 2023+338 and XTE J1118+480, and $\ga$4 times larger than 
the accretion disks. Therefore, the mid-infrared emission is produced in a
region that extends beyond the binary orbit.

Jets could produce an emitting region larger than the binary orbit, but 
given the spectra of the mid-infrared excesses we suggest that jets are 
only minor contributors. Two LMXBs in our sample 
have recently been detected as radio sources in quiescence: GS 2023+338 with  
flat-spectrum radio emission with an intensity of 350$\mu$Jy over the 
frequency range 1.4--8.4 GHz \citep{gal05}, and A 0620$-$00 with a 
radio flux of 50 $\mu$Jy at 8.4 GHz \citep{gal06}.
The 24$\mu$m excesses from A 0620$-$00 is equal to the flux in the radio,
so it could be produced by a jet with a flat ($S_\nu \propto \nu^\alpha$, 
with $\alpha \approx 0$) spectrum
between the radio and mid-infrared. For GS 2023+338, the 24 $\mu$m flux
is slightly lower than the radio flux, which would imply that its jet
has a steeper $\alpha \ga -0.3$ spectrum.  However, the 8 $\mu$m excesses 
from A 0620$-$00, GS 2023+338, and XTE J1118+480 are not consistent with a 
flat-spectrum radio jet, because for XTE J1118+480 and A 0262$-$00 they are 
4 times larger than the excesses at 24 $\mu$m, and for GS 2023+338 it is 10 
times larger. If we assume the 24 $\mu$m emission is from a 
flat-spectrum jet, then it contributes $<$ 25\% to the 
8 $\mu$m flux from XTE J1118+480 and A 0620$-$00, and $<$10\% from 
GS 2023+338.

We suggest that the excesses from 
A 0620$-$00 and XTE J1118+480 originate
from circumbinary dust that re-processes the light
of the companions. To estimate the masses of the dust, we 
assume it is contained in optically thick disks. The disks could lie as 
close to the center of masses of the binaries
as 1.7$a$, at which point they would be tidally truncated 
\citep{taam03,dub04}. We can estimate
the temperature profile of such disks by assuming they are flat, in 
which case
\begin{equation}
T_{\rm disk} \approx \left( \frac{2}{3\pi}\right)^{1/4} \left( \frac{R_*}{R_{\rm disk}} \right)^{3/4} T_*,
\end{equation}
where $T_{\rm disk}$ and $R_{\rm disk}$ are the temperatures and radii of 
the disks, and $T_*$ and $R_*$ are the temperatures and radii of the 
companions \citep[e.g.,][]{rp91,jur03}.
Using Equation~\ref{eq:orbit_sep} and the parameters in 
Table~\ref{tab:targets}, we would expect the inner edges of the circumbinary
disks to have temperatures of $\approx$600 K, which is consistent
with our detection of excesses that peak near 8 $\mu$m. The
lack of excess flux at 24$\mu$m implies that material does not re-process 
much stellar light beyond radii $\approx$3 times larger than the binary 
separations. If we assume
the disks are composed of dust with an opacity 
$\chi_\nu$$\sim$300 cm$^{2}$ g$^{-1}$ at 8 $\mu$m \citep[e.g.,][]{dl84}, 
then for them to have an optical depth of $\tau$$\sim$1, the disks would 
only need to contain $\sim$$10^{22}$ g of dust. If we assume the emitting
material is optically thin, we find dust masses that are similar within 
a factor of a few \citep[e.g., Eq. 3 in][]{eva97}.
If the gas-to-dust ratio is similar to that of the interstellar medium, 
$\sim$100, then the total mass of the circumbinary material could be 
$\sim$$10^{24}$ g.

\section{Conclusions}

The detection of excess mid-infrared flux from A 0620$-$00 and XTE J1118+480
(Fig.~\ref{fig:sed}) 
provides evidence that circumbinary material is present around some
LMXBs, but its spectrum suggests that it will only be detectable in the 
mid-infrared. The excess emission peaks at 8 $\mu$m, and if we model it as a 
single-temperature blackbody, we predict that the circumbinary material
produced negligible flux shortward of $\approx$3 $\mu$m. Moreover,
the lack excess mid-infrared emission from Cen X-4 demonstrates
that such material is not ubiquitous. Finally, for GS 2023+338, the 
mid-infrared emission appears to originate from a hot accretion disk, 
and the contribution of the accretion disk to the mid-infrared light 
can be predicted based on the optical measurements (the dotted line in 
Fig.~\ref{fig:sed}). These results provide
reassurance that efforts to determine the inclinations of LMXBs by 
modeling the ellipsoidal modulations of their infrared and optical light
curves are not compromised by the presence of circumbinary material.

For A 0620$-$00 and XTE J1118+480, we suggest that the circumbinary 
disks are either the remains of fall-back
disks produced in the supernovae that formed the compact objects 
\citep[e.g.,][]{wck06,cs06}, or material injected into circumbinary
orbits during the process of mass loss by the Roche-lobe filling 
companions \citep[e.g.,][]{dub04}. These circumbinary disks contain 
$\sim$$10^{22}$ g of dust, which for a standard dust-to-mass ratio of 
$\sim$100 implies a total mass of $\sim$$10^{24}$ g. 
If the disks 
are produced by matter ejected into circumbinary orbits during the 
process of mass transfer, then it represents only a tiny fraction of the
$\sim$$10^{32}$ g ($\sim$0.1 \msun) that will be lost by the companions over 
the lifetimes of these LMXBs. Alternatively, \citet{cs06} have proposed 
that fall-back disks containing 
$\ga$$10^{24}$ g of material could form asteroids and inject them into
the magnetospheres of pulsars with a high enough rate to explain the 
the observed intermittency in their radio pulses. Asteroids might not 
survive the outbursts of LMXBs \citep{mh01}, but the mid-infrared excesses 
that we have identified could be the remnants of similar fall-back disks.

However, the circumbinary material is not massive enough 
either to affect the evolution of the orbital angular momentum of
LMXBs \citep[e.g., $\sim$$10^{29}$ g in][]{taam03}, 
or to form planetesimals \citep[e.g., $\sim$$10^{28}$ g in][]{mh01}. 
As the formation and evolution of circumbinary matter around LMXBs and CVs 
are considered further, it may be that the paucity of matter around 
A 0620$-$00 and XTE J1118+480 (and the apparent lack of circumbinary matter 
around the one neutron star LMXB in our sample, Cen X-4) may be the most 
lasting aspect of this result.

\acknowledgements
We are grateful to T. Barman for providing the model stellar spectra,
to E. Gallo and S. Migliari for alerting us to a mistake in our 
original photometry, to M. Jura for conversations about circumstellar dust,
to J. McClintock for discussions about the contribution of the accretion 
disk to the mid-infrared light, and to
the referee for comments that clarified the text. 
Support for this work was provided by NASA through an award issued by 
JPL/Caltech.

\end{document}